\documentstyle[sprocl]{article}
\def\Journal#1#2#3#4{{#1} {\bf #2}, #3 (#4)}
\def\NPB{{\em Nucl. Phys.} B}
\def\PLB{{\em Phys. Lett.}  B}

\def\be{\begin{equation}}
\def\ee{\end{equation}}
\def\bea{\begin{eqnarray}}
\def\eea{\end{eqnarray}}

\begin{document}
\begin{flushright}
NORDITA-96/65 P\\
hep-ph/9610212
\end{flushright}
\title{
THEORETICAL UNCERTAINTIES IN THE DETERMINATION OF $\alpha_s$
FROM HADRONIC EVENT OBSERVABLES: THE IMPACT OF NONPERTURBATIVE
EFFECTS
      }
\author{V.M. BRAUN }
\address{NORDITA, Blegdamsvej 17, DK--2100 Copenhagen \O,
Denmark }

\bigskip

\address{talk presented at Annual Meeting of Division of Particles
        and Fields of the APS (DPF`96), Minneapolis, MN, August 10--15. 1996} 
\maketitle\abstracts{
%Precision measurements of $\alpha_s$ from the structure of hadronic 
%final states (jet rates, event shapes etc.) require understanding of
%the structure of nonperturbative power-suppressed corrections 
%in situations when the Operator Product Expansion is not applicable.
%I give a short review of the present status of this problem, report on
%some recent progress, and speculate on possibilities for further 
%development.
I review the theoretical status of hadronization corrections to hadronic
event observables and discuss their impact on the determination of $\alpha_s$.
                    }
%\section{Introduction}

Precise determination of the strong coupling has become one of the most
important tasks of the QCD theory and phenomenology \cite{alpha}. 
From the theoretical
point of view the cleanest measurement comes from the total $e^+e^-$
annihilation cross section at high energies. 
The accuracy is dominated in this case by statistical errors; high 
statistics is needed because the effect proportional to $\alpha_s$
is a few percent fraction only of the total cross section.
Thus, going over to various event shape observables which are proportional
to $\alpha_s$ at the leading order presents a clear experimental advantage.
The price to pay is that the QCD description becomes more
complicated and making a defensible estimate of the theoretical
accuracy presents a nontrivial task.

The uncertainty due to nonperturbative
effects is a particularly delicate issue.
 From the experimentalist's point of view this is the 
uncertainty of `hadronization corrections' which are applied 
to uncover the 
structure of the event at the {\it parton level} from the observed 
structure at the {\it hadron level}. 
%It is tacitly assumed that 
%the partonic structure of the event is perturbatively calculable
%(resummation of certain classes of contributions may be necessary in parts
%of the phase space), while the transition to hadrons 
%is 'truly' nonperturbative.
The hadronization process is  modelled in a certain way, 
and differences between models 
(say, Lund string fragmentation or parton showers) are 
taken to estimate the error.
 In statistically
important regions the hadronization corrections are of order
10\% (at the Z peak) while the claimed error is of order 2--3\%.
This is less than the uncertainties of  
existing perturbative calculations
(estimated by the scale dependence) so that for an experimentalist
the hadronization could be considered  under control.

This procedure is very successful phenomenologically,
 but it is unsatisfactory
from the theoretical point of view. In particular 
--- I will return to this point later --- 
the separation of perturbative versus
nonperturbative alias  parton level versus hadronization effects 
is theoretically ill-defined, and the commonly accepted procedures 
might be suspected to be plagued by double counting of 
infrared effects. The theoretical understanding 
is guided by the Wilson operator product expansion (OPE)
which is applicable, most notably, to the 
deep inelastic scattering (DIS).  
As a representative example,
consider the Gross--Llewellyn Smith sum rule (GLS):
\begin{equation}
\int_0^1 \!\!dx\, F_3^{\nu p +\bar\nu p}(x,Q^2)
=3\Bigg[1-\frac{\alpha_s}{\pi}(Q^2)+\ldots -\frac{8}{27}
\frac{\langle\!\langle O\rangle\!\rangle}{Q^2}\Bigg]+O(1/Q^4)
\end{equation} 
Here $\langle\!\langle O\rangle\!\rangle$ is the reduced matrix element
of a certain quark-antiquark-gluon operator which quantifies the
correlations between partons in the nucleon.

From the OPE one learns:
(i)~$\underline{Power~counting}$ of nonperturbative effects; in this case that
    perturbation theory is valid to $O(1/Q^2)$ accuracy and that the
    description to order $O(1/Q^2)$ requires one dimensionful 
    nonperturbative parameter;
(ii)~`$\underline{Universality}$' of nonperturbative effects in the sense 
     that the same quark-antiquark-gluon operator appears in different
     physical processes; the coefficient is calculable in perturbation
     theory, hence one can sacrifice one measurement to get the prediction
     for  other ones.

Theoretical approaches to the nonperturbative corrections
in hadron production would aim to get the similar structure.
For a generic observable dominated by short distances
one expects an expansion of the type
\begin{equation}\label{powerexpand}
R = \sum_{n=0}^\infty r_n \alpha_s^n + \left(\frac{\Lambda_R}{Q}\right)^p
\left(\ln Q/\Lambda\right)^\gamma
\end{equation}
The goal is to understand the power counting, that is which powers $p$
are present, calculate `anomalous dimensions' $\gamma$ and 
relate the nonperturbative parameters $\Lambda_R$ in different processes
(universality). Note that the question of actual magnitude of the
nonperturbative parameters remains open.

For the  precision $\alpha_s$ measurements the 
power counting is of primary importance. Indeed, assuming a $1/Q^2$ 
correction, for the
effective hard scale $Q_{\rm eff}\ge 10$ GeV and the intrinsic size
of  `matrix elements' 1 GeV, one obtains a
ball-park estimate for the nonperturbative effects $\sim$ 1\%
which can be neglected. On the other hand, if the 
nonperturbative effects are  $O(1/Q)$,
they can be of order 10\% and one has to estimate
them quantitatively.

The structure of nonperturbative
power-suppressed corrections to physical observables appears to be
 intimately connected with the same-sign factorial divergence $r_n\sim n!$ 
of the QCD perturbation theory in high orders.
Divergence of perturbation theory implies that
 the sum of the series is only defined to power accuracy
  $\exp[-{\rm const}/\alpha_s(Q)] \sim 1 /Q^p$
  and the ambiguity has to be remedied by adding 
nonperturbative corrections.
In particular,
the large-n  behavior of the perturbative 
coefficients
\begin{equation}
  r_n \sim {\rm const}\cdot n^{\tilde\gamma}\cdot s^{-n}\cdot n!,
\end{equation}
necessarily requires  a nonperturbative correction
in (\ref{powerexpand})
with the powers $p$ and $\gamma$ related in a simple way
to the coefficients $s$ and $\tilde\gamma$, respectively.

This has two consequences. First,
one can investigate the 
structure of nonperturbative corrections to a large class of observables
by studying the structure of higher orders in the perturbative expansions,
which  attracted a lot of recent activity (see \cite{B} for a recent review).
 Here is
a short summary of results, related to $\alpha_s$ determinations:
\begin{itemize}
 \item{} Most of the existing hadronic event observables are predicted 
         to have nonperturbative corrections of order $1/Q$ \cite{W,DW,AZ}. 
 \item{} The thrust and heavy jet mass distributions 
         have $1/Q$ corrections for the average values. However,
         outside the two-jet region nonperturbative effects have 
         and extra suppression factor $\sim\alpha_s(Q)$ \cite{NS}.        
 \item{} The one-particle inclusive cross section in the $e^+ e^-$ 
         annihilation (fragmentation function) has only $1/Q^2$ 
         corrections for fixed energy fraction;
          however, the integrated longitudinal (and transverse)
         cross sections have $1/Q$ corrections \cite{DW2,BBM}.
 \item{} Power counting of the nonperturbative effects in jet
         fractions depends on the jet finding algorithm; $1/Q$ corrections
         are intrinsic for JADE and are most likely absent for the
         Durham $k_\perp$-clustering method \cite{DMW}.
 \item{} There are no $1/Q$ corrections to the Drell-Yan (heavy quark 
         production) cross section to leading order \cite{BBDY};
         their existence to the $O(\alpha_s(Q)/Q)$ 
         accuracy is still disputed \cite{critics}.
\end{itemize}

The power counting of nonperturbative effects can be tested 
experimentally by the energy dependence of
hadron event observables, subtracting the parton level prediction.
This was done recently by DELPHI \cite{DELPHI}. 
%The results tend to
%agree to the theoretical expectations, although this agreement 
%is not conclusive.

Perturbative calculations in
certain regions of phase space may require resummations of large
logarithms (threshold corrections).
An important question is whether the 
power counting of nonperturbative effects is disturbed by 
the resummations. It was studied for the Drell-Yan production
 \cite{CS,KS,BBDY,BBtalk}. 
Different resummation 
procedures which are equivalent in perturbation theory,
can introduce different power-like corrections, and it was suggested that
a criterium for a `good' resummation technique is that it does not bring in
nonperturbative effects which are absent in finite orders; this study 
emphasizes importance of large-angle soft gluon emission \cite{BBDY,BBtalk}.
It turns out that quality of resummations in moment and
momentum spaces is not related in any obvious way and their truncation
may introduce spurious $1/Q^p$ effects with a small power $p \sim \alpha_s$
\cite{sudakon}.  

The second  consequence 
 is that the separation of `perturbative' and `nonperturbative'
effects is conceptually ambiguous. To make it 
meaningful one has to  introduce an {\it IR matching scale 
$\mu_{\rm IR}$} \cite{OPE},
define `nonperturbative'
contributions as absorbing all effects from scales below $\mu_{\rm IR}$ and
subtract the small momentum contributions from perturbative series 
to avoid a double counting. Continuing the example with
the GLS sum rule one obtains, schematically
\begin{equation}
\int_0^1 \!\!dx\, F_3^{\nu p +\bar\nu p}(x,Q^2)
=3\Bigg[1-\Bigg(1-\frac{\mu^2_{\rm IR}}{Q^2}\Bigg)
\frac{\alpha_s}{\pi}(Q^2)+\ldots -\frac{8}{27}
\frac{\langle\!\langle O^{(\mu_{\rm IR})}\rangle\!\rangle}{Q^2}\Bigg]+O(1/Q^4)
\end{equation} 
%Note that coefficients of the perturbative expansion are corrected 
%by power suppressed terms, and the matrix element 
% $\langle\!\langle O^{(\mu_{\rm IR})}\rangle\!\rangle$
%depends on the IR matching scale.
The premium for this refinement is that  perturbation theory 
restricted to the contributions of scales above $\mu_{\rm IR}$
is (almost) free from factorial divergences in high orders; the price to
pay is that subtraction of small momenta is  very awkward 
in practice and introduces
an additional scale (and scheme) dependence.

The same applies to  the `hadronization corrections'. 
The lessons to be learnt from OPE are: (i) separation of the parton cascade
 and hadronization is ambiguous; (ii) high orders of the 
perturbative series can imitate a power correction; (iii) if extracted
from comparison with the data, the power correction is expected to depend 
on the order of perturbation theory, factorization scheme and scale;
(iv) numerical estimates suggest that `true nonperturbative' 
 contributions at small scales are of the same order as perturbative.

To see how it works, consider energy dependence of the 
mean value of thrust $\langle 1-T\rangle$ in the $e^+e^-$ annihilation.
The experimental data are well described by the formula
\begin{equation}
\langle 1-T\rangle(Q) =0.335\,\alpha_s(Q)+1.02\,\alpha_s^2(Q)+
 \frac{1 \mbox{\rm GeV}}{Q}
\end{equation}
Here $Q$ is the c.m. energy, the two first terms on the r.h.s.
correspond to the perturbation theory and the last 
term is the nonperturbative correction.
The common procedure is to fix the 1~GeV/Q correction (assume it
comes from a certain model) and fit $\alpha_s(Q)$ to the data; 
this gives $\alpha_s(M_Z) = 0.120$. Then, {\it keeping the 
hadronization correction fixed}, one varies the scale $\alpha_s(Q)\to
\alpha_s(\mu)$ to estimate the perturbative uncertainty coming from
unknown higher orders.

The caveat is that if  
nonperturbative and higher-order perturbative effects are inseparable, 
the hadronization correction  
can be scale-dependent itself, an aspect which usually remains fogged.
It should be fitted
anew for each scale. Let us try $\mu=0.13\, Q$ which is of order of 
the gluon transverse momentum $k_\perp^2\sim(1-T)Q^2/4$.
Keeping  
$\alpha_s(M_Z)=0.120$ fixed, one obtains an equally good fit to the
data by changing the $1/Q$ correction 
\begin{equation}
\langle 1-T\rangle(Q) =0.335\,\alpha_s(0.13 Q)+0.19\,\alpha_s^2(0.13Q)+
 \frac{0.4 \mbox{\rm GeV}}{Q}
\end{equation}
This illustrates that the reshuffling of higher-order perturbative corrections
by changing the factorization scale can be compensated by
the change of the hadronization correction.
The only known way to separate the nonperturbative effects from
the factorization scale dependence is to 
introduce an IR matching scale in the spirit of the OPE 
treatment as discussed above.
This is attempted in the Dokshitser-Webber model \cite{DW},
where contributions of small scales are  subtracted from 
perturbation theory.
First  applications of this model to the extraction of $\alpha_s$
are encouraging \cite{DELPHI}: The scale dependence is reduced by factor 
two compared to the traditional treatment, the IR matching scale 
 dependence is small, and the value of the extracted nonperturbative 
parameter is stable.

The question of possible `universality' of 
hadronization corrections is complicated and needs further study.
Real solution requires development of the OPE-like techniques.
The present approaches to this problem rely on two assumptions: 
that nonperturbative effects are proportional to perturbative ambiguities and
that present `naive' estimates of these ambiguities are representative. 
This may well be not true, or only partially true, and these assumptions 
should be tested on simpler examples. Of particular interest are 
predictions for the x-dependence of the $1/Q^2$ effects in DIS 
\cite{DMW,M,DW1} and
for the fragmentation functions \cite{DW2,BBM}.
 The Dokshitser-Webber model 
has a buit-in universality also for the event shapes. 

In general, one should expect that the expansion
is organized in powers of some physical scale (say the BLM scale)
rather than the `naive' c.m. energy, and differences in the 
effective scales for various processes (together with the 
power counting) can give a rough idea 
of the difference in nonperturbative corrections.
For example, the twist-4 effects in DIS at $x\rightarrow 1$ are 
proportional to
$\Lambda^2/[(1-x)Q^2]$ while they are of order 
$\Lambda^2/[(1-x)^2Q^2]$ for the Drell-Yan cross section.
The difference reflects different hard scales for the gluon emission.

Viewed this way, the problem of power corrections is inseparable from the
familiar problem of the scale dependence and scale fixing. 
The specifics of hadron event
observables is that the physical scale (estimated e.g. by any of existing
scale fixing prescriptions) appears to be very low -- 
 of order 1/10 of the c.m. energy. 
It was argued \cite{burrows} that the data do not show any 
preference for a low scale since the spread of the results for $\alpha_s$ 
between different observables
is not reduced. From my point of view this 
conclusion is not warranted and the observed spread may indicate 
an intrinsic accuracy of the present treatment, with the separation 
of perturbative and hadronization effects as independent entities.
\footnote{
Using low scales is more consistent with
hadronization corrections taken from parton shower models
which imply $\alpha_s(k_\perp)$ for each gluon emission.
Note, however, that these corrections by construction parametrize
 contributions of small gluon virtualities, which
is in spirit of the OPE. Combining them
with the fixed-order perturbative calculations 
one has to subtract contributions of the low scales
from the latter to avoid the double counting,  at least in principle.
In practice this is difficult to do because the parton showers 
do not include NLO radiative corrections consistently.}

In any case, some standartization is needed for the choice of scale
in the data analysis. At present, each experimental collaboration uses
different criteria to fix the scale range,
 which makes the comparison difficult.
%To give an example, the value of $\alpha_s(M_Z)$ extracted from
%the hadronic event observables with the resummation of 
%threshold corrections is larger than without resummation
%in the LEP summary \cite{...} while the effect is opposite   
% in the SLD analysis \cite{SLD}. The difference is (most likely)  in the 
%different range of scales. 

To summarize,  what is done and what is necessary to do  to improve the
theoretical accuracy of the determinations of $\alpha_s$ from hadronic
event shapes?
\begin{itemize}
\item{} Certain theoretical tools are developed to determine the 
        power counting of nonperturbative effects in hadronic event
        observables. Some of the existing event shapes are predicted to
        have smaller corrections than the others, and it is advisable
        to concentrate on them rather than make `global fits'.
\item{} With this new knowledge, one should try  to design new event 
       shapes. An `event shape  of the year' should be measurable, calculable
        to $O(\alpha_s^3)$ accuracy and have small nonperturbative effects.
        Some work in this direction is reported in \cite{BBM}.
\item{} In addition to the traditional procedures, one has to apply
        alternative methods for the
       data analysis with the hadronization corrections fitted to the data 
       rather  than taken from models. 
       One should always bear in mind that
       hadronization corrections do not have objective meaning unless a clear 
       scale separation is made. 
\item{}The existing procedures should be checked
       for double counting of perturbative contributions at low scales.
       The complete next-to-leading order parton shower models would help a
       lot.
\item{} A convention for the scheme and scale setting is badly needed. 
\end{itemize}
My personal feeling is that present theoretical accuracy of  $\alpha_s(M_Z)$ 
 from hadronic  event shapes is of order 7\% 
and it can be improved up to factor
two if the above questions are clarified.

\section*{Acknowledgments} I am grateful to M. Beneke, K. Hamacher
and  L. L\"onnblad
for the discussions. Special thanks are
are due to M. Shifman for the invitation to this conference.


\section*{References}
\begin{thebibliography}{99} 
\bibitem{alpha} 
S. Bethke, presented at `QCD-96', Montpellier, July 1996 \\
 (hep-ex/9609014); 
M.~Schmelling, presented  at ICHEP`96, Warsaw, July 1996. 
\bibitem{B} M. Beneke, presented at ICHEP`96, Warsaw, July 1996\\
              (hep-ph/9609215).
\bibitem{W} B. Webber, \Journal{\PLB}{339}{148}{1994}.
\bibitem{DW} Yu. Dokshitser and B. Webber,
              \Journal{\PLB}{352}{451}{1995}.
\bibitem{AZ} R. Akhoury and V.I. Zakharov, \Journal{\PLB}{357}{646}{1995}.
\Journal{\NPB}{465}{295}{1996}.
\bibitem{NS} P. Nason and M.H. Seimour, 
             \Journal{\NPB}{454}{291}{1995}.
\bibitem{DW2} M. Dasgupta and B. Webber, CAVENDISH-HEP-96-2 (hep-ph/9608394). 
\bibitem{BBM} M. Beneke, V.M. Braun and L. Magnea, NORDITA-96/52 P\\
 (hep-ph/9609266).
\bibitem{DMW} Yu. Dokshitser, G. Marchesini and B. Webber,
 CERN-TH-95-281\\ (hep-ph/9512336).
\bibitem{BBDY} M. Beneke and V.M. Braun, \Journal{\NPB}{454}{253}{1995} 
\bibitem{critics} G. Korchemsky, presented at NORDITA workshop on renormalons
and power corrections in QCD, Copenhagen, August 1996; R. Akhoury, ibid.
\bibitem{DELPHI} DELPHI collab., DELPHI 96-107 CONF 34, paper pa02003
 submitted to ICHEP'96, Warsaw.
\bibitem{CS} H. Contopanagos and G. Sterman, \Journal{\NPB}{419}{77}{1994}.
\bibitem{KS} G.P. Korchemsky and G. Sterman, \Journal{\NPB}{437}{415}{1995}.
\bibitem{BBtalk} M. Beneke and V.M. Braun, presented at  
   `Continuos Advances in QCD' Minneapolis, March 1996 (hep-ph/9605375).
%\bibitem{SLD} SLD Collab., K. Abe {\it et al},
%              \Journal{\PRD}{51}{962}{1995}.
\bibitem{sudakon} S. Catani et al., CERN-TH-96-86 (hep-ph/9604351).
\bibitem{OPE} V. Novikov {\it et al}, \Journal{\NPB}{249}{445}{1985}.
\bibitem{M} E. Stein {\it et al}, \Journal{\PLB}{376}{177}{1996}; 
          M. Meyer-Hermann {\it et al}, UFTP-414-1996 [hep-ph/9605229].
\bibitem{DW1} M. Dasgupta and B. Webber, CAVENDISH-HEP-96-1 (hep-ph/9604388). 
\bibitem{burrows} P.N. Burrows {\it et al},
   \Journal{\PLB}{382}{157}{1996}.
\end{thebibliography}
\end{document}